# Clustering Unstructured Data (Flat Files)
An Implementation in Text Mining Tool

Yasir Safeer[1], Atika Mustafa[2] and Anis Noor Ali[3]
Department of Computer Science
FAST – National University of Computer and Emerging Sciences
Karachi, Pakistan
[1]yasirsafeer@gmail.com, [2]atika.mustafa@nu.edu.pk, [3]anisnoorali@hotmail.com

*Abstract*—With the advancement of technology and reduced storage costs, individuals and organizations are tending towards the usage of electronic media for storing textual information and documents. It is time consuming for readers to retrieve relevant information from unstructured document collection. It is easier and less time consuming to find documents from a large collection when the collection is ordered or classified by group or category. The problem of finding best such grouping is still there. This paper discusses the implementation of k-Means clustering algorithm for clustering unstructured text documents that we implemented, beginning with the representation of unstructured text and reaching the resulting set of clusters. Based on the analysis of resulting clusters for a sample set of documents, we have also proposed a technique to represent documents that can further improve the clustering result.

## I. INTRODUCTION

Text Mining uses unstructured textual information and examines it in attempt to discover structure and implicit meanings "hidden" within the text [6]. Text mining concerns looking for patterns in unstructured text [7].

A cluster is a group of related documents, and clustering, also called unsupervised learning is the operation of grouping documents on the basis of some similarity measure, automatically without having to pre-specify categories [8]. We do not have any training data to create a classifier that has learned to group documents. Without any prior knowledge of number of groups, group size, and the type of documents, the problem of clustering appears challenging [1].

Given N documents, the clustering algorithm finds *k*, number of clusters and associates each text document to the cluster. The problem of clustering involves identifying number of clusters and assigning each document to one of the clusters such that the intra-documents similarity is maximum compared to inter-cluster similarity.

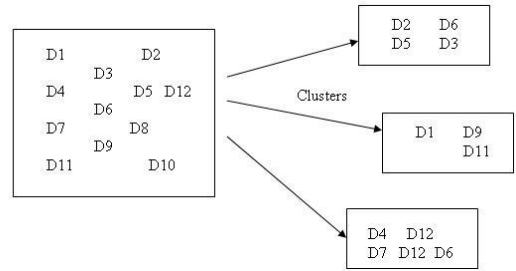

Figure 1. Document Clustering

One of the main purposes of clustering documents is to quickly locate relevant documents [1]. In the best case, the clusters relate to a goal that is similar to one that would be attempted with the extra effort of manual label assignment. In that case, the label is an answer to a useful question. For example, if a company is operating at a call center where users of their products submit problems, hoping to get a resolution of their difficulties, the queries are problem statements submitted as text. Surely, the company would like to know about the types of problems that are being submitted. Clustering can help us understand the types of problems submitted [1]. There is a lot of interest in the research of genes and proteins using public databases. Some tools capture the interaction between cells, molecules and proteins, and others extract biological facts from articles. Thousands of these facts can be analyzed for similarities and relationships [1]. Domain of the input documents used in the analysis of our implementation, discussed in the following sections, is restricted to Computer Science (CS).

## II. REPRESENTATION OF UNSTRUCTURED TEXT

Before clustering algorithm is used, it is necessary to give structure to the unstructured textual document. The document is represented in the form of vector such that the words (also called features) represent dimensions of the vector and frequency of the word in document is the magnitude of the vector. i.e.

- A Vector is of the form
  $<(t_1,f_1),(t_2,f_2),(t_3,f_3),\ldots,(t_n,f_n)>$

where $t_1,t_2,..,t_n$ are the terms/words(dimension of the vector) and $f_1,f_2,\ldots,f_n$ are the corresponding frequencies or magnitude of the vector components.

A few tokens with their frequencies found in the vector of the document [9] are given below:

TABLE I. LIST OF FEW TOKENS WITH THEIR FREQUENCY IN A DOCUMENT

| Tokens | Freq. | Tokens | Freq. |
|---|---|---|---|
| oracle | 77 | cryptographer | 6 |
| attacker | 62 | terminate | 5 |
| cryptosystem | 62 | turing | 5 |
| problem | 59 | return | 5 |
| function | 52 | study | 5 |
| key | 46 | bit | 4 |
| secure | 38 | communication | 4 |
| encryption | 27 | service | 3 |
| query | 18 | k-bit | 3 |
| cryptology | 16 | plaintext | 3 |
| asymmetric | 16 | discrete | 2 |
| cryptography | 16 | connected | 2 |
| block | 15 | asymptotic | 2 |
| cryptographic | 14 | fact | 2 |
| decryption | 12 | heuristically | 2 |
| symmetric | 12 | attacked | 2 |
| compute | 11 | electronic | 1 |
| advance | 10 | identifier | 1 |
| user | 8 | signed | 1 |
| reduction | 8 | implementing | 1 |
| standard | 7 | solvable | 1 |
| polynomial-time | 7 | prime | 1 |
| code | 6 | computable | 1 |
| digital | 6 | | |

The algorithm of creating a document vector is given below [2]:

TABLE II. GENERATING FEATURES FROM TOKENS

**Input**
Token Stream (TS), all the tokens in the document collection

**Output**
HS, a Hash Table of tokens with respective frequencies

**Initialize:**
Hash Table (HS):= empty Hash Table

**for each** Token in Token Stream (TS) **do**
    **If** Hash Table (HS) contains Token **then**
      Frequency:= value of Token in hs
      increment Frequency by 1
    **else**
      Frequency:=1
    **enidif**
    store Frequency as value of Token in Hash Table (HS)
**endfor**

output HS

Creating a dimension for every unique word will not be productive and will result in a vector with large number of dimensions of which not every dimension is significant in clustering. This will result in a synonym being treated as a different dimension which will reduce the accuracy while computing similarity. In order to avoid this problem, a Domain Dictionary is used which contains most of the words of Computer Science domain that are of importance. These words are organized in the form of hierarchy in which every word belongs to some category. The category in turn may belong to some other category with the exception of root level category.

Parent-category → Subcategory→Subcategory → Term(word). e.g. Databases→RDBMS→ERD→Entity

Before preparing vector for a document, the following techniques are applied on the input text.
- The noise words or stop words are excluded during the process of Tokenization.
- Stemming is performed in order to treat different forms of a word as a single feature. This is done by implementing a rule based algorithm for Inflectional Stemming [2]. This reduces the size of the vector as more than one forms of a word are mapped to a single dimension.

The following table [2] lists dictionary reduction techniques from which Local Dictionary, Stop Words and Inflectional Stemming are used.

TABLE III. DICTIONARY REDUCTION TECHNIQUES

| Local Dictionary |
|---|
| Stop Words |
| Frequent Words |
| Feature Selection |
| Token Reduction: Stemming, Synonyms |

*A. tf-idf Formulation And Normalization of Vector*

To achieve better predictive accuracy, additional transformations have been implemented to the vector representation by using *tf-idf* formulation. The *tf-idf* formulation is used to compute weights or scores of a word. In (1), the weight $w(j)$ assigned to word $j$ in a document is the *tf-idf* formulation, where $j$ is the $j$-th word, $tf(j)$ is the frequency of word $j$ in the document, $N$ is the number of documents in the collection, and $df(j)$ is the number of documents in which word $j$ appears.

Eq. (1) is called inverse document frequency (*idf*). If a word appears in many documents, its *idf* will be less compared to the word which appears in a few documents and is unique. The actual weight of a word, therefore, increases or decreases depending on *idf* and is not dependent on the term frequency alone. Because documents are of variable length, frequency information could be misleading. The *tf-idf* measure can be normalized to a unit length of a document D as described by *norm(D)* in (3) [2]. Equation (5) gives the cosine distance.

$$w(j) = tf(j) * \log_2(N/df(j)) \quad (1)$$

$$\log_2(N/df(j)) = idf(j) \quad (2)$$

$$norm(D) = \sqrt{\sum w(j)^2} \quad (3)$$

$$\hat{D} = \vec{D}/norm(D) \quad (4)$$

$$cosine(d1, d2) = \sum (w_{d1}(j) * w_{d2}(j))/(norm(d1) * norm(d2)) \quad (5)$$

e.g.
For three vectors (after removing stop-words and performing stemming),

- *Doc1 < (computer, 60), (JAVA, 30)...>*
- *Doc2 < (computer, 55), (PASCAL, 20)...>*
- *Doc3 < (graphic, 24), (Database, 99)...>*

Total Documents, *N*=3

The vectors shown above indicate that the term '*computer*' is less important compared to other terms (such as '*JAVA*' which appears in only one document out of three) for identifying groups or clusters because this term appears in more number of documents (two out of three in this case) making it less distinguishable feature for clustering. Whatever the actual frequency of the term may be, some weight must be assigned to each term depending on the importance in the given set of documents. The method used in our implementation is the *tf-idf* formulation.

In *tf-idf* formulation the frequency of term *i*, tf(*i*) is multiplied by a factor calculated using inverse-document-frequency *idf(i)* given in (2). In the example above, total number of documents is *N*=3, the term frequency of '*computer*' is $tf_{computer}$ and the number of documents in which the term '*computer*' occurs is $df_{computer}$. For Doc1,

$$idf_{computer} = \log_2(N/df_{computer})$$
$$= \log_2(3/2)$$
$$= 0.5849$$

*tf-idf* weight for term '*computer*' is,
$$w_{computer} = tf_{computer} * idf_{computer}$$
$$= 60 * 0.5849$$
$$= 35.094$$

Similarly,
$$idf_{JAVA} = \log_2(N/df_{JAVA})$$
$$= \log_2(3/1)$$
$$= 1.5849$$
$$w_{JAVA} = tf_{JAVA} * idf_{JAVA}$$
$$= 30 * 1.5849$$
$$= 47.547$$

After *tf-idf* measure, more weight is given to '*JAVA*' (the distinguishing term) and the weight of '*computer*' is much less (since it appears in more documents), although their actual frequencies depict an entirely different picture in the vector of *Doc1* above. The vector in *tf-idf* formulation can then be normalized using (4) to obtain the unit vector of the document.

### III. MEASURING SIMILARITY

The most important factor in a clustering algorithm is the similarity measure [8]. In order to find the similarity of two vectors, *Cosine similarity* is used. For cosine similarity, the two vectors are multiplied, assuming they are normalized [2]. For any two vectors *v1, v2* normalized using (4),

Cosine *Similarity* (*v1, v2*) =

< (a1, c1), (a2, c2)...> . <(x1, k1), (x2, k2), (x3, k3)...>
= (c1) (k1) + (c2) (k2) + (c3) (k3) + ...

where '.' is the ordinary dot product (a scalar value).

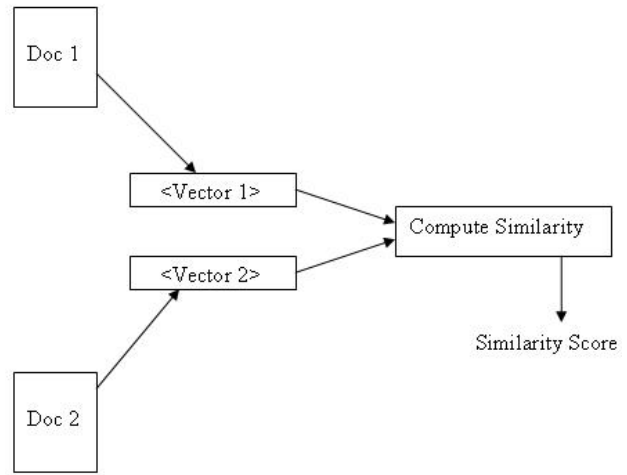

Figure 2. Computing Similarity

## IV. REPRESENTING A CLUSTER

The cluster is represented by taking average of all the constituent document vectors in the cluster. This results in a new summarized vector. This vector, like other vectors can be compared with other vectors, therefore, comparison between *document-document* and *document-cluster* follows the same method discussed in section III.

For cluster '*c*' containing two documents,
- *v1 < (a, p1), (b, p2)...>*
- *v2 < (a, q1), (b, q2)...>*

cluster representation is merely a matter of taking vector average of the constituent vectors and representing it as a *composite document* [2]. i.e. a vector as the average (or mean) of constituent vectors

*Cluster {v1, v2} = < (p1+q1)/2, (p2+q2)/2...>*

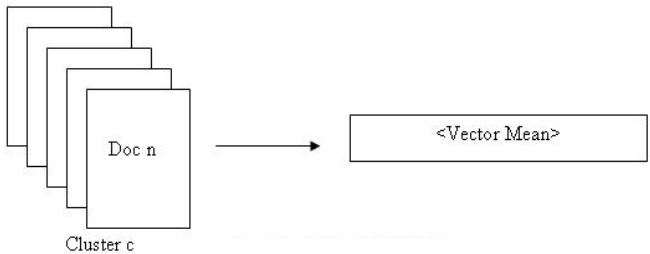

Figure 3. Cluster Representation

## V. CLUSTERING ALGORITHM

The algorithm that we implemented is *k*-Means clustering algorithm. This algorithm takes *k*, number of initial bins as parameter and performs clustering. The algorithm is provided below [2]:

TABLE IV. THE *K*-MEANS CLUSTERING ALGORITHM

1. Distribute all documents among k bins.
   *A bin is an initial set of documents that is used before the algorithm starts. It can also be considered as initial cluster.*
   a. The mean vector of the vectors of all documents is computed and is referred to as 'global vector'.
   b. The similarity of each document with the global vector is computed.
   c. The documents are sorted on the basis of similarity computed in part b.
   d. The documents are evenly distributed to k bins.
2. Compute mean vector for each bin.
   *As discussed in section IV.*
3. Compare the vector of each document to the bin means and note the mean vector that is most similar.
   *As discussed in section III.*
4. Move all documents to their most similar bins.
5. If no document has been moved to a new bin, then stop; else go to step 2.

## VI. DETERMINING *K*, NUMBER OF CLUSTERS

*k*-Means algorithm takes *k*, number of bins as input, therefore the value of *k* cannot be determined in advance without analyzing the documents. *k* can be determined by first performing clustering for all possible cluster size and then selecting the *k* that gives the minimum total variance, *E(k)* (error) of documents with their respective clusters. Note that the value of *k* in our case ranges from 2 to N. Clustering with *k=1* is not desired as single cluster will be of no use. For all the values of *k* in the given range, clustering is performed and variance of each result is computed as follows [2]:

$$E(k) = \sum_{i=1}^{n} \frac{(x^i - m_{ci})^2}{n}$$

where $x^i$ is the *i-th* document vector, $m_{ci}$ is its cluster mean and $c_i \in \{1,....,k\}$ is its corresponding cluster index.

Once the value of *k* is determined, each cluster can be assigned a label by using categorization algorithm [2].

## VII. CLUSTERING RESULT

An input sample of 24 documents [11, 12, 13, 14, 15, 16, 17, 18, 19, 20, 21, 22, 23, 24, 25, 26, 27, 28, 29, 30, 31, 32, 33, 34] were provided to the *k*-Means Algorithm. With the initial value of *k*=24, the algorithm was run for three different scenarios:

(a) When the document vectors were formed on the basis of features (words) of the document.

(b) When the document vectors were formed on the basis of sub-category of features.

(c) When the document vectors were formed on the basis of parent category of the feature.

The result of *k*-means clustering algorithm for each case is given below:

TABLE V. CLUSTERS- ON THE BASIS OF FEATURE VECTORS

| Cluster name | Documents |
|---|---|
| Text Mining | [12, 13, 14, 16, 17, 18, 28] |
| Databases | [11, 25, 26, 27] |
| Operating Systems | [23, 32] |
| Mobile Computing | [22, 24] |
| Microprocessors | [33, 34] |
| Programming | [30, 31] |
| Data Structures | [29] |
| Business Computing | [20, 21] |
| World Wide Web | [15] |
| Data Transfer | [19] |

TABLE VI. CLUSTERS – ON THE BASIS OF SUBCATEGORY VECTORS[4]

| Cluster name | Documents |
|---|---|
| Text Mining | [12, 13, 14, 16, 17, 18, 28, 31] |
| Databases | [11, 25, 26, 27] |
| Operating Systems | [21, 23, 32] |
| Communication | [22, 24] |
| Microprocessors | [33, 34] |
| Programming Languages | [30] |
| Data Structures | [29] |
| Hardware | [20] |
| World Wide Web | [15] |
| Data Transfer | [19] |

TABLE VII. CLUSTERS – ON THE BASIS OF PARENT CATEGORY VECTORS[4]

| Cluster name | Documents |
|---|---|
| Software | [11, 12, 14, 16, 17, 25, 26, 27, 28, 30] |
| Operating Systems | [22, 23, 24] |
| Hardware | [31, 32, 33, 34] |
| Text Mining | [13, 18] |
| Network | [19, 20, 21 29] |
| World Wide Web | [15] |

---

[4] the decision of selecting parent category vectors or sub-category vectors depends on the total number of root (parent) level categories, levels of sub-categories and organization of the domain dictionary used. A better, rich and well organized domain dictionary directly affects document representation; yields better clustering result and produces more relevant cluster names.

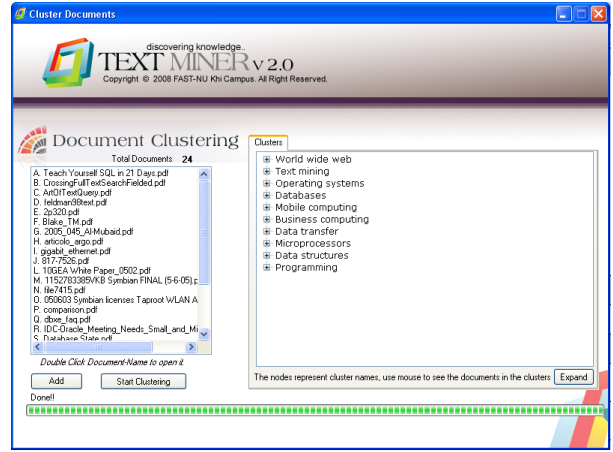

Figure 4. Document Clustering in our Text Mining Tool

## VIII. TECHNIQUE FOR IMPROVING THE QUALITY OF CLUSTERS

### A. Using Domain Dictionary to form vectors on the basis of sub-category and parent category

The quality of clusters can be improved by utilizing the domain dictionary which contains words in a hierarchical fashion.

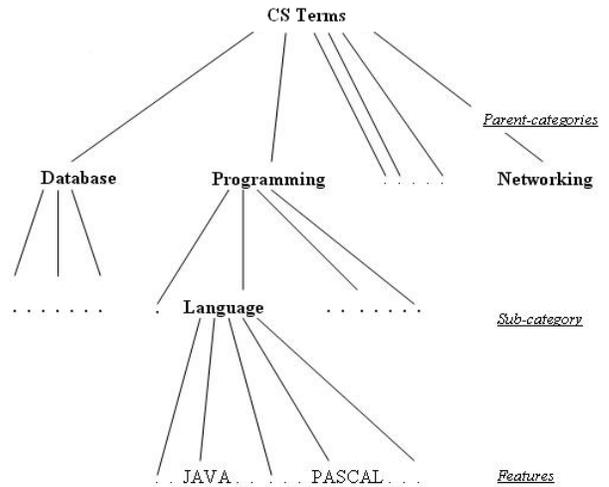

Figure 5. Domain Dictionary (CS Domain)

For every word w and sub-category s,

$$w \; R \; s \quad (6)$$

iff w comes under the sub-category s in the domain dictionary, where R is a binary relation.

The sub-category vector representation of a document with features,
$< (w_1, f_{w1}), (w_2, f_{w2}) ... (w_n, f_{wn}) >$
is
$< (s_1, f_{s1}), (s_2, f_{s2}) ... (s_m, f_{sm}) >$
where *n* is the total number of unique features (words) in the document.

$$\forall n\ \exists m \ni w_n\ R\ s_m$$

for some 1<=*m*<=*c* (c is total number of sub-categories)
$f_{wn}$ is the frequency of the *n-th* word
$f_{sm}$ is the frequency of *m-th* sub-category
R is defined in (6)

Consider a *feature* vector with features (words in a document) as vector dimension:

Document1< (register, 400), (JAVA, 22)... >

The *sub- category* vector of the same document is:

Document1<(architecture, 400+$K_1$), (language, 22+$K_2$)..>

where $K_1$ and $K_2$ are the total frequencies of other features that come under the sub-category '*architecture*' and '*language*' respectively.

Sub-category and parent category vectors generalize the representation of document and the result of document similarity is improved.

Consider two documents that are written on the topic of '*Programming language*', both documents are similar in nature but the difference is that one document is written on programming JAVA and the other on programming PASCAL. If document vectors are made on the basis of *features*, both the documents will be considered less similar because not both the documents will have the term 'JAVA' or 'PASCAL' (even though both documents are similar as both come under the category of *programming* and should be grouped in same cluster).

If the same documents are represented on the basis of sub-category vectors then regardless of whether the term JAVA occurs or PASCAL, the vector dimension used for both the terms will be '*programming language*' because both 'JAVA' and 'PASCAL' come under the sub-category of '*programming language*' in the *domain dictionary*. The similarity of the two documents will be greater in this case which improves the quality of the clusters.

## IX. FUTURE WORK

So far our work is based on predictive methods using frequencies and rules. The quality of result can be improved further by adding English Language semantics that contribute in the formation of vectors. This will require incorporating some NLP techniques such as POS tagging (using Hidden Markov Models, HMM) and then using the tagged terms to determine the importance of features. A tagger finds the most likely POS tag for a word in text. POS taggers report precision rates of 90% or higher [10]. POS tagging is often part of a higher-level application such as Information Extraction, a summarizer, or a Q&A system [1]. The importance of the feature will not only depend on the frequency itself, but also on the context where it is used in the text as determined by the POS tagger.

## X. CONCLUSION

In this paper we have discussed the concept of document clustering. We have also presented the implementation of k-means clustering algorithm as implemented by us. We have compared three different ways of representing a document and suggested how an organized domain dictionary can be used to achieve better similarity results of the documents. The implementation discussed in this paper is limited only to predictive methods based on frequency of terms occurring in the document, however, the area of document clustering needs to be further explored using language semantics and context of terms. This could further improve similarity measure of documents which would ultimately provide better clusters for a given set of documents.

AUTHORS PROFILE

**Yasir Safeer** received his BS degree in Computer Science from FAST - National University of Computer and Emerging Sciences, Karachi, Pakistan in 2008. He was also awarded Gold Medal for securing 1st position in BS in addition to various merit based scholarships during college and undergraduate studies. He is currently working as a Software Engineer in a software house. His research interests include text mining & information extraction and knowledge discovery.

**Atika Mustafa** received her MS and BS degrees in Computer Science from University of Saarland, Saarbruecken, Germany in 2002 and University of Karachi, Pakistan in 1996 respectively. She is currently an Assistant Professor in the Department of Computer Science, National University of Computer and Emerging Sciences, Karachi, Pakistan. Her research interests include text mining & information extraction, computer graphics(rendering of natural phenomena, visual perception).

**Anis Noor Ali** received his BS degree in Computer Science from FAST - National University of Computer and Emerging Sciences, Karachi, Pakistan in 2008. He is currently working in an IT company as a Senior Software Engineer. His research interests include algorithms and network security.